\documentclass[twocolumn]{article}

\usepackage[dvips]{epsfig}
\usepackage[]{amsbsy}

\def\thebibliography#1{\section*{\normalsize \bf References 
 }\list
 {[\arabic{enumi}]}{\settowidth\labelwidth{[#1]}\leftmargin\labelwidth
 \advance\leftmargin\labelsep
 \usecounter{enumi}}
 \def\newblock{\hskip .11em plus .33em minus .07em}
 \sloppy\clubpenalty4000\widowpenalty4000
 \sfcode`\.=1000\relax}

\begin{document}

\twocolumn[

\begin{center} \large \bf
   Surface metal-insulator transition in the Hubbard model
\end{center}
\vspace{-3mm}

\begin{center} 
   M. Potthoff and W. Nolting
\end{center}
\vspace{-6mm}

\begin{center} \small \it 
   Institut f\"ur Physik, 
   Humboldt-Universit\"at zu Berlin, 
   Germany
\end{center}
\vspace{2mm}

\begin{center}
\parbox{141mm}{ \small
The correlation-driven metal-insulator (Mott) transition at a solid
surface is studied within the Hubbard model for a semi-infinite 
lattice by means of the dynamical mean-field theory. The transition 
takes place at a unique critical strength of the interaction. 
Depending on the surface geometry, the interaction strength and the
wave vector, we find one-electron excitations in the coherent part 
of the surface-projected metallic spectrum which are confined to two 
dimensions. 
}
\end{center}
\vspace{8mm} 
]

% PACS:

% 71.10.Fd Lattice fermion models (Hubbard model, etc.)
% 71.27.+a Strongly correlated electron systems; heavy fermions
% 71.30.+h Metal-insulator transitions 
%          and other electronic transitions
% 73.20.-r Surface and interface electron states

{\center \bf \noindent I. INTRODUCTION \\ \mbox{} \\} 

The correlation-driven transition from a paramagnetic metal to a
paramagnetic insulator is a fundamental problem in condensed-matter 
physics. The Mott transition is of interest since strong electron 
correlations lead to low-energy electronic properties that are 
qualitatively different from those predicted by band theory.

Theoretical attempts started with the early ideas of Mott 
\cite{Mot61}, Hubbard \cite{Hub64b} and Brinkman and Rice 
\cite{BR70} (for an overview cf.\ \cite{Geb97}). In recent years 
a more detailed understanding has been achieved by studying the 
half-filled Hubbard model in the limit of high spatial dimensions 
$d$ \cite{MV89}. Here the electronic self-energy is a local 
quantity, and a self-consistent mapping onto the single-impurity 
Anderson model (SIAM) becomes possible \cite{GK92a,Jar92}. The 
impurity model can be solved numerically 
\cite{Jar92,RZK92,GK92b,CK94,SRKR94}
or, for weak or strong coupling, by using perturbative approaches 
\cite{GK92a,OPK97}.
This constitutes the dynamical mean-field theory (DMFT) which is 
exact for $d=\infty$ but remains a powerful method also at finite 
$d$ \cite{Vol93,GKKR96}.

Extensive DMFT studies for the $d=\infty$ Bethe lattice with a
semi-elliptic free density of states have established the 
following scenario for the Mott transition \cite{GKKR96}: 
For strong interaction $U$ the two high-energy charge excitation 
peaks are well separated by an insulating gap in the one-electron 
spectrum similar to the Hubbard-III approach \cite{Hub64b}. 
For decreasing $U$ the insulating solution ceases to exist at a 
critical value $U_{c1}$. On the other hand, for small $U$ the 
system is a metallic Fermi liquid with a quasi-particle band at 
the Fermi energy. For increasing $U$ the metallic solution of the 
DMFT equations coalesces with the insulating one at another critical
interaction $U_{c2}>U_{c1}$. As $U$ approaches $U_{c2}$ the 
effective mass diverges as in the Brinkman-Rice solution 
\cite{BR70}. Between $U_{c1}$ and $U_{c2}$ there is a 
region where both solutions coexist. 
For $T\ne 0$ the insulator is stable below
$U_{c2}$, and the transition is of first order \cite{GKKR96}.
For $T=0$ entropic effects favor the metallic solution. 
The transition is thus expected \cite{GKKR96,RMK94,MSKR95,
BPH98} to take place at $U_{c2}$ (second-order transition) or at 
least close to $U_{c2}$. 

Although some interesting questions are yet unsolved (e.~g.\ 
concerning the
concept of a {\em pre-formed gap} for $U \mapsto U_{c2}$
\cite{Keh98}), one can state that 
there is a comparatively detailed understanding of the Mott 
transition in infinite dimensions. For finite dimensions one
has to assume that the mean-field picture provided by the 
DMFT remains valid qualitatively. This is the basis for the
present study which considers the Mott transition at $T=0$, 
but concentrates on a new aspect of the problem: 
the solid surface. Surface effects are of interest for different 
reasons:

i) Taking into account the surface comes closer to the (inverse)
photoemission experiment which determines the spectral function.
The information depth in low-energy electron spectroscopy is usually 
limited to a few surface layers only \cite{Pow88}. 

ii) The breakdown of translational symmetry due to the surface 
introduces a layer dependence in the physical quantities which is 
worth studying. The layer dependence of the quasi-particle weight 
is calculated within the DMFT and also within conventional 
second-order perturbation theory 
around the Hartree-Fock solution (SOPT-HF) \cite{PN97c}. 
The perturbative treatment applies to the weak-coupling regime,
non-local terms can be studied here. The (non-perturbative) DMFT
accounts for the local (temporal) fluctuations exactly, but neglects 
all spatial correlations. Within both approaches it is found that 
the presence of the surface gives rise to strong oscillations 
depending on $U$ and geometry.

iii) One may ask whether or not the critical interaction strength
is modified at the surface, analogously to a (possibly) enhanced 
Curie temperature at the surface of a ferromagnet \cite{Mil71}. 
For uniform model parameters we find a unique critical interaction 
$U_{c2}$ where the whole system undergoes the transition. This 
excludes a surface phase different from the bulk.

iv) The low-energy electronic structure is of particular interest
close to the critical point. Here but also well below $U_{c2}$, 
the surface renormalization of the one-electron 
excitation energies is shown to be sufficiently 
strong to generate a surface mode. The mode is essentially 
confined to two dimensions and shows up with a reduced dispersion. 
The generating mechanism for this type of surface excitation is 
exclusively due to electron correlations. A simple criterion for 
their existence is derived analytically.
\\

{\center \bf \noindent II. MODEL AND GREEN FUNCTION \\ \mbox{} \\} 

We study the Hubbard model at half-filling and $T=0$:
\begin{equation}
  H = \sum_{ij\sigma} t_{ij} 
  c^\dagger_{i\sigma} c_{j\sigma}
  + \frac{U}{2} \sum_{i\sigma} n_{i\sigma} n_{i-\sigma} \: ,
\label{eq:hubbard}
\end{equation}
with $i$ and $j$ running over the sites of a semi-infinite $d=3$ 
lattice which is chosen here to be the simple cubic lattice cut at 
a low-index lattice plane. The uniform nearest-neighbor hopping 
$t_{\langle ij \rangle} \equiv -t$ with $t=1$ fixes the energy 
scale. To focus on the Mott transition from a paramagnetic metal to a
paramagnetic insulator, we restrict ourselves to the spin-symmetric
solutions of the DMFT equations as usual \cite{Geb97}. Thereby, we 
set aside the fact that on a bipartite lattice the half-filled 
Hubbard model is actually in an antiferromagnetic phase below a 
finite N\'eel temperature.
 
We concentrate on the metallic spectrum 
in the vicinity of the Fermi energy. This is governed by the 
low-energy expansion of the self-energy \cite{Lut61}: 
\begin{equation}
  \Sigma_{ij}(E) = \delta_{ij} \frac{U}{2} + \beta_{ij} E + 
  i \gamma_{ij} E^2 + \cdots \: .
\label{eq:sigmaexp}
\end{equation} 
With the help of the (symmetric) matrices $\boldsymbol{\beta}
= (\beta_{ij})$ and ${\bf Z}=({\bf 1} - \boldsymbol{\beta})^{-1}$ 
the one-electron Green function 
$G_{ij}(E) = \langle\langle c_{i\sigma} ; c_{j\sigma}^\dagger 
\rangle \rangle$ can be written in the form:
\begin{equation}
  {\bf G}(E) = {\bf Z}^{\frac{1}{2}} \frac{\bf 1}{E {\bf 1} -
  {\bf Z}^{\frac{1}{2}} {\bf T} {\bf Z}^{\frac{1}{2}}} 
  {\bf Z}^{\frac{1}{2}} \: ,
\label{eq:greencoh}
\end{equation}
where ${\bf G}=(G_{ij})$ and ${\bf T}=(t_{ij})$. Only the term 
linear in $E$ is taken into account. In particular, we consider 
an energy range where damping effects ($\propto i\gamma_{ij} E^2$)
are unimportant.

This {\em coherent} part of the 
spectrum is exclusively determined by the so-called
quasi-particle weight matrix ${\bf Z}=(z_{ij})$ 
(mass-enhancement matrix ${\bf Z}^{-1}$). The usual {\em bulk} 
quasi-particle weight $z({\bf q})$ is obtained from $z_{ij}$
by three-dimensional Fourier transformation. For the 
{\em semi-infinite} system only lateral translational symmetry 
can be exploited. Considering the system 
to be built up by $N_\perp$ layers parallel to the surface 
($N_\perp \mapsto \infty$), there is a collection of $N_\perp$ 
one-dimensional Fermi-``surfaces'' in the two-dimensional surface 
Brillouin zone (SBZ). The $\mu$-th Fermi surface is given by
$\epsilon_{{\bf k}\mu}=0$ where $\epsilon_{{\bf k}\mu}$ 
($\mu=1,...,N_\perp$) are the eigenvalues of the
renormalized hopping matrix 
${\bf Z}^{\frac{1}{2}} {\bf T} {\bf Z}^{\frac{1}{2}}$ at a 
two-dimensional wave vector $\bf k$ of the SBZ.
The corresponding eigenvectors ${\bf u}_{{\bf k}\mu}$
yield the discontinuous drops of the momentum distribution function
along the direction $\bf k$ in the SBZ via:
$\delta n({\bf k} = {\bf k}_{\rm F}^\mu) = N_\perp^{-1} 
{\bf u}_{{\bf k}\mu}^\dagger \cdot {\bf Z} \cdot 
{\bf u}_{{\bf k}\mu}$. 
For $N_\perp \mapsto \infty$ the decrease of $n({\bf k})$ is 
continuous, the quasi-particle weight matrix
$\bf Z$, however, has a well-defined meaning according to Eq.
(\ref{eq:greencoh}).
\\

{\center \bf \noindent III. PERTURBATIONAL APPROACH \\ \mbox{} \\} 

$\bf Z$ shall be calculated at a mean-field level assuming that spin 
and charge fluctuations are reasonably {\em local}. This implies a 
local self-energy, $\Sigma_{ij} \approx \delta_{ij} \Sigma_i$, which 
needs justification for $d<\infty$ and in particular for the $d=3$ 
semi-infinite lattice. Some insight can be gained by conventional 
second-order perturbation theory around the 
Hartree-Fock solution. SOPT-HF is capable of accounting
for the complete non-locality of the self-energy in the 
weak-coupling regime also for the case of reduced translational 
symmetry \cite{PN97c}. 

%++++++++++++++++++++++++++++++++++++++++++++++++++++++++++
\begin{figure}[b] 
\vspace{-4mm}
\psfig{figure=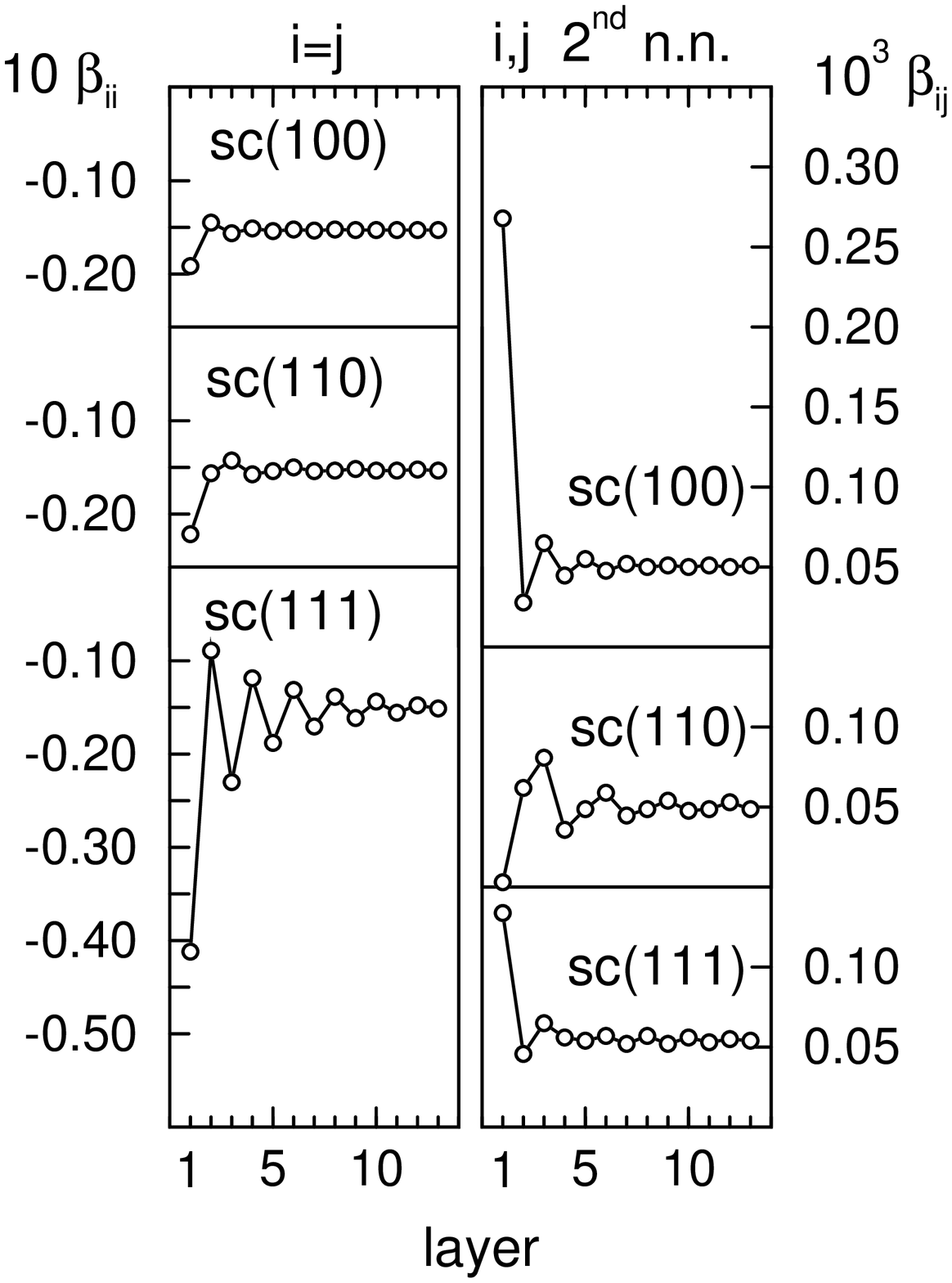,width=80mm,angle=0}
\vspace{-2mm}

\parbox[]{85mm}{\small Fig.~1.
Layer dependence of the on-site (left) and the off-site (right)
linear coefficient in the low-energy expansion (\ref{eq:sigmaexp}) 
of the self-energy at different surfaces of the s.c. lattice as 
obtained within SOPT-HF. Second-nearest neighbors $i$ and $j$ within 
the same layer. Within SOPT-HF $\beta_{ij} \sim U^2 \times 
\mbox{const.}$; we set $U=1$ in the figure. Note the different 
scales. Energy units such that $t=1$.
\label{fig:betalayer}
}
\end{figure}
%++++++++++++++++++++++++++++++++++++++++++++++++++++++++++

%++++++++++++++++++++++++++++++++++++++++++++++++++++++++++
\begin{figure}[t] 
\vspace{3mm}
\centerline{\psfig{figure=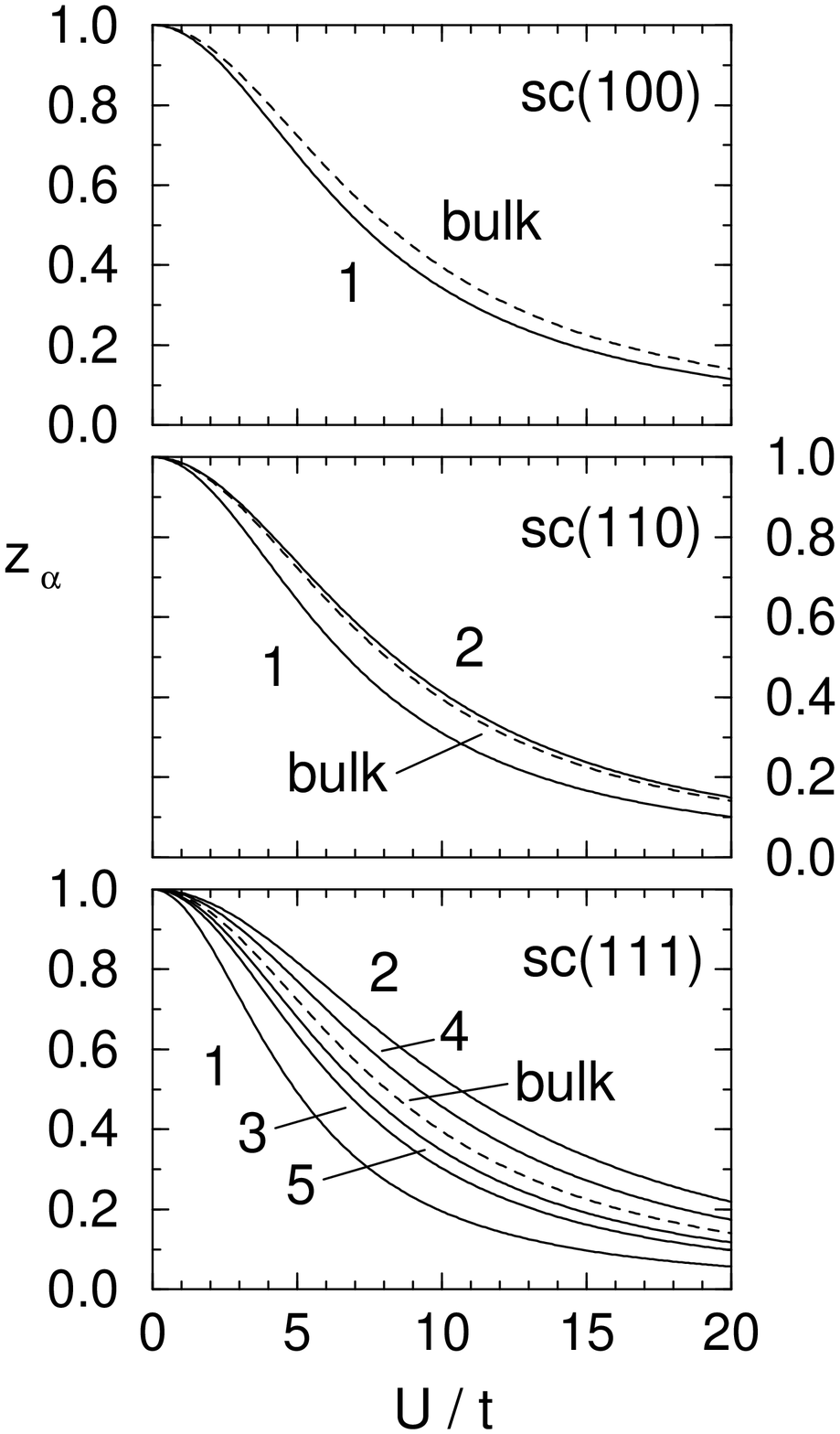,width=65mm,angle=0}}
\vspace{3mm}

\parbox[]{85mm}{\small Fig.~2.
$U$ dependence of the layer-dependent quasi-particle weight 
$z_\alpha = (1 - \beta_\alpha)^{-1}$ within SOPT-HF. Results 
for different surfaces. `1' stands for the topmost surface 
layer. Dashed lines: bulk s.c. lattice.
\label{fig:zlayer}
}
\end{figure}
%++++++++++++++++++++++++++++++++++++++++++++++++++++++++++

We have calculated the linear coefficient $\beta_{ij}$ for the 
different low-index surfaces. The low-energy expansion of the
SOPT-HF self-energy yields:
\begin{equation}
  \beta_{ij} = -2 U^2 
  \int_0^\infty \!\!\!\! dx 
  \int_0^\infty \!\!\!\! dy 
  \int_0^\infty \!\!\!\! dz \:
  \frac{\rho^{(0)}_{ij}(x) \rho^{(0)}_{ij}(y) \rho^{(0)}_{ij}(z)}
  {(x+y+z)^2} \; ,
\label{eq:beta}
\end{equation}
if the HF (on- or off-site) density of states is symmetric,
$\rho^{(0)}_{ij}(-E)=\rho^{(0)}_{ij}(E)$, and $\beta_{ij}=0$ if it 
is antisymmetric. Fig.~1 shows the local term $\beta_{i=j}$ to exhibit 
a weak layer dependence for the sc(100) surface. Only for the 
topmost surface layer a significant difference to the bulk value 
is found. Surface effects become stronger for the sc(110) and are 
most pronounced for the sc(111) surface where a strong oscillation 
of $\beta_{ii}$ is noticed. This can be understood as follows: 
According to Eq.\ (\ref{eq:beta}) the dominant contribution to 
$\beta_{ii}$ comes from the HF density of states density near $E=0$. 
For the sc(111) surface, however, the $E=0$ value is known \cite{KS71} 
to oscillate between zero (for the sub-surface and all even layers) 
and a constant value (for the topmost and all odd layers). This 
anomalous behavior is found \cite{KS71} to dominate the shape of the 
density of states within an energy range $\Delta E \sim t L^{-2}$, 
where $L$ denotes the distance to the surface. Passing from the surface 
to the volume, $L \mapsto \infty$, the bulk HF density of states is 
recovered outside an infinitesimally small energy range around $E=0$. 
For the coefficient $\beta_{ii}$ this implies a damped layer-by-layer 
oscillation with the maximum absolute value for the topmost layer.

For the discussion of the non-local terms $\beta_{i\ne j}$ we have
to consider second nearest-neighbors since electron-hole symmetry
requires $\rho^{(0)}_{ij}(E)$ to be antisymmetric and thus 
$\beta_{ij} = 0$ for nearest neighbors 
$i,j$ on a bipartite lattice and at half-filling. We also find 
surface-induced oscillations in the case $i\ne j$ as can be seen
in Fig.~1. More important, 
however, the absolute values are smaller by about 
{\em two orders of magnitude} compared with the local terms.

Let us consider the quasi-particle weight matrix.
Decomposing the linear coefficient into local and non-local parts,
$\beta_{ij} = \delta_{ij} \beta_{ij} + (1 - \delta_{ij}) \beta_{ij}$,
and expanding ${\bf Z}=({\bf 1} - \boldsymbol{\beta})^{-1}$ in powers 
of the non-local one, yields 
\begin{equation}
  \Delta z_{ii} = \sum_j^{j\ne i} 
  \frac{1}{1 - \beta_{ii}} \beta_{ij} 
  \frac{1}{1 - \beta_{jj}} \beta_{ji}
  \frac{1}{1 - \beta_{ii}}
\end{equation}
for the lowest-order non-local correction $\Delta z_{ii}$ of 
the local element $z_{ii}$. The correction is of the order 
$(\beta_{i\ne j})^2$ and can thus be neglected in the 
weak-coupling limit.

Fig.~2 shows the local elements of the quasi-particle weight 
matrix $z_{ii} \approx 1/(1 - \beta_{ii})$ within SOPT-HF as 
a function of $U$. For weak coupling we observe the quadratic
dependence on the interaction strength: $1-z_\alpha(U) \propto U^2$. 
Depending on geometry, there is a considerable layer dependence 
even for weak interaction. For strong $U$, SOPT-HF still predicts 
a Fermi-liquid state with $z_\alpha >0$. Here, as expected, the 
perturbational approach breaks down. 
\\

{\center \bf \noindent IV. DYNAMICAL MEAN-FIELD THEORY \\ \mbox{} \\} 

We now turn to the dynamical mean-field theory to consider the 
intermediate- and strong-coupling regime. DMFT is a non-perturbative 
method. However, we have to {\em assume} that a local self-energy 
functional is a reasonable approximation for $d=3$ dimensions. 
From the perturbational results, this is well justified
in the weak-coupling regime, for larger $U$ the assumption
may be questioned. In any case, we expect DMFT to be a good 
starting point.

We have generalized the mean-field equations to the case of film 
geometries. A film consisting of a sufficiently large but finite 
number of layers $N_\perp$ is considered to simulate the actual 
surface. Apart from mirror symmetry at the film center, the layers 
are treated as being inequivalent. DMFT must thus be applied in the 
following way: We start with a guess for the (local) self-energy 
$\Sigma_\alpha(E)$ for each layer (translational symmetry is assumed
{\em within} the layers). Second, the Dyson equation is solved on 
the semi-infinite lattice to get the on-site Green functions 
$G_{\alpha}(E)$. For this purpose we make use of lateral 
translational symmetry which restricts the numerical calculations
to inversion of $N_\perp \times N_\perp$ matrices on a wave-vector 
mesh in the two-dimensional SBZ. Corresponding to each layer 
$\alpha$, an effective impurity problem is set up. From the on-site 
Green function and the self-energy, the $\alpha$-th SIAM is 
specified by fixing the respective free impurity Green function 
$G_{\rm imp}^{(0)}(E)$ via the DMFT self-consistency condition:
\begin{equation}
  G_{\rm imp}^{(0)}(E) = 
  \left( G_{\alpha}(E)^{-1} + \Sigma_\alpha(E) \right)^{-1} \: .
\label{eq:dmft}
\end{equation}
The crucial step is the solution of the ($\alpha$-th) SIAM to get 
the self-energy $\Sigma_{\alpha}(E)$ which is required for the 
next cycle. Applying DMFT to a film geometry implies a
self-consistent treatment of $N_\perp$ impurity problems that are 
coupled indirectly via the respective baths of conduction electrons.

For finite temperatures the impurity problems can be solved by 
employing the Quantum-Monte-Carlo method \cite{Jar92,RZK92,GK92b} 
using the Hirsch-Fye algorithm. For $T=0$ the Exact-Diagonalization 
(ED) approach of Caffarel and Krauth \cite{CK94,SRKR94} may be 
applied and will be chosen for the present study. ED is able to 
yield the essentially exact solution of the mean-field equations 
in a parameter range where the errors introduced by the finite 
system size are unimportant. For the Mott problem the relevant 
low-energy scale is set by the width of the quasi-particle peak 
in the metallic solution. It has to be expected that there are 
non-negligible finite-size effects when this energy scale becomes 
comparatively small. We are thus limited to interactions strengths 
that are not too close to $U_{c2}$ and cannot access the very 
critical regime. 

The algorithm proceeds as follows: The DMFT self-consistency 
condition yields a bath Green function $G_{\rm imp}^{(0)}(iE_n)$. 
The parameters of a SIAM with $n_s$ sites, the conduction-band 
energies $\epsilon_k$ and the hybridization strengths $V_k$ 
($k=2,...,n_s$), are obtained by minimizing the following cost 
function:
\begin{equation}
  \chi^2 = 
  \frac{1}{n_{\rm max}+1} 
  \sum_{n=0}^{n_{\rm max}}
  \left|
  G_{\rm imp}^{(0)}(iE_n)^{-1} - G_{\rm n_s}^{(0)}(iE_n)^{-1}
  \right| \: ,
\label{eq:chi2}
\end{equation}
where $G_{\rm n_s}^{(0)}(iE_n - \mu) = (iE_n - \epsilon_d - 
\sum_k V_{k}^2/(iE_n - \epsilon_{k}))^{-1}$ is the free ($U=0$) 
Green function of the $n_s$-site SIAM. The fit of 
$G_{\rm imp}^{(0)}(iE_n)^{-1}$ is performed on the imaginary axis, 
$iE_n= i(2n+1)\pi/\widetilde{\beta}$ with a {\em fictitious} inverse
temperature $\widetilde{\beta}$ which introduces a low-energy 
cutoff. Lancz\`o's technique \cite{Hay80} is used to calculate the 
zero-temperature impurity Green function $G_{\rm imp}(iE_n)$. The 
local self-energy of the $\alpha$-th layer is then obtained via 
$\Sigma_\alpha(iE_n)=G_{\rm imp}^{(0)}(iE_n)^{-1} 
- G_{\rm imp}(iE_n)^{-1}$. At half-filling electron-hole symmetry 
can be enforced by $\epsilon_{k}=-\epsilon_{k'}$ and 
$V^2_{k}=V^2_{k'}$ with $k+k'=n_s+2$.
\\

{\center \bf \noindent V. RESULTS AND DISCUSSION \\ \mbox{} \\} 

Routinely, the calculations have been performed for $n_s=8$ sites. 
For interaction strengths well below $U_{c2}$ this has turned out to 
be sufficient for convergence. The results are independent of the 
high-energy cutoff $n_{\rm max}$. A small low-energy cutoff 
($\widetilde{\beta}W \sim 10^3$) is necessary to obtain a converged 
value for $1 - 1/z_\alpha = (\partial/\partial(iE)) \mbox{Im} 
\Sigma_\alpha(iE=0)$. Although the film geometry implies a 
high-dimensional parameter space ($N_\perp(n_s-1)/2$, i.~e.\ 
$\approx 100$ parameters to be determined self-consistently) we 
always obtained a stabilized metallic or insulating solution.
A moderate number of layers $N_\perp$ in the film is sufficient 
to simulate the semi-infinite system. We used $N_\perp = 11,15,21$ 
layers to investigate the (100), (110) and (111) surfaces, 
respectively. The convergence has been checked by comparing 
the results from calculations for different $N_\perp$.

%++++++++++++++++++++++++++++++++++++++++++++++++++++++++++
\begin{figure}[t] 
\vspace{-4mm}
\centerline{\psfig{figure=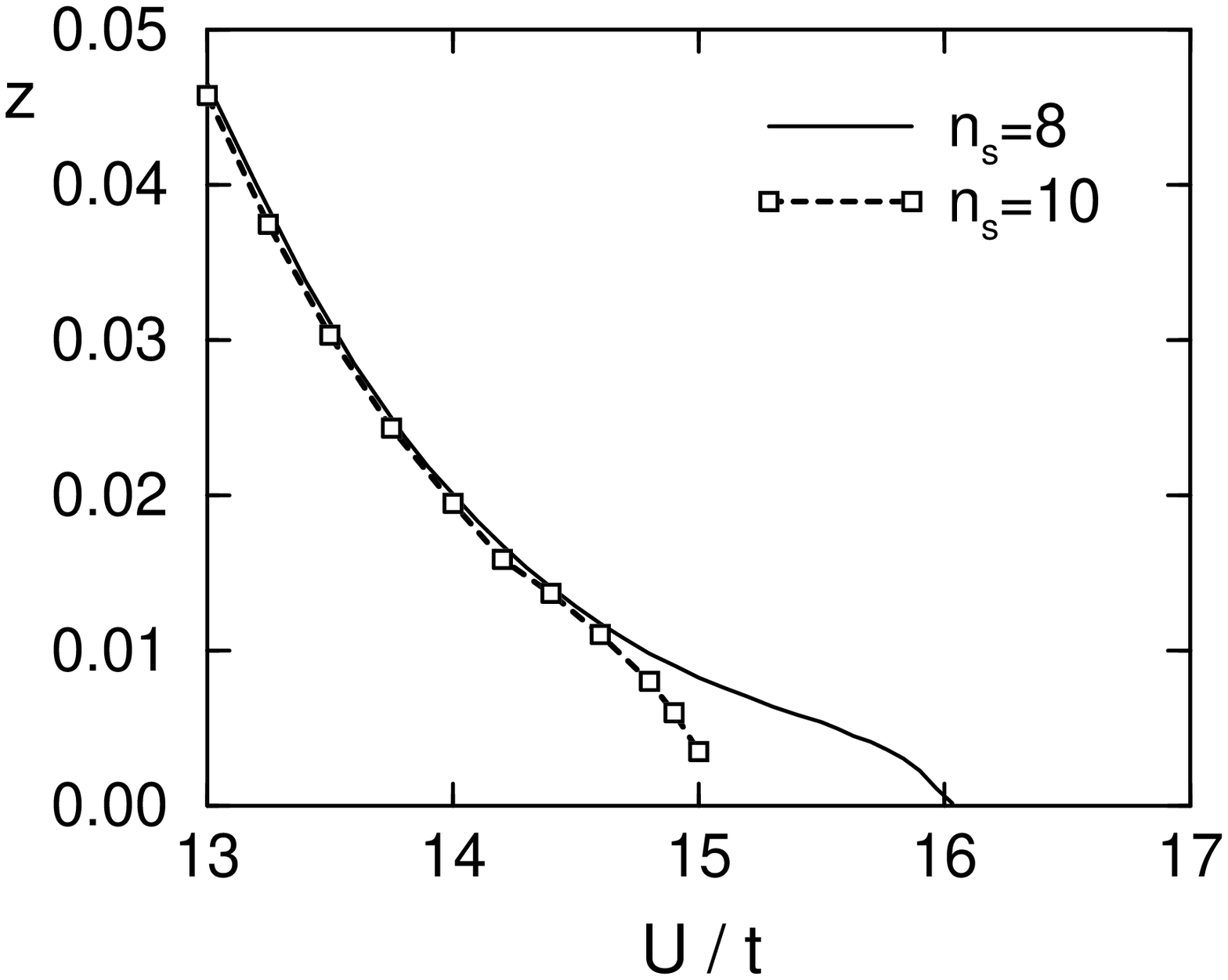,width=80mm,angle=0}}
\vspace{-4mm}

\parbox[]{85mm}{\small Fig.~3.
Bulk quasi-particle weight as a function of $U$. Results for
$n_s=8$ and $n_s=10$.
\label{fig:ns}
}
\end{figure}
%++++++++++++++++++++++++++++++++++++++++++++++++++++++++++

Difficulties arise for $U\mapsto U_{c2}$.
Due to the finite number of sites considered, the energy 
resolution of the conduction band cannot be better than 
$\Delta E \approx \eta W / n_s$, where $W$ is the free band 
width and $\eta$ is a constant which accounts for the fact 
that the conduction-band energies $\epsilon_k$ are not equally 
spaced and depend on $U$. $\Delta E$ can be estimated by comparing
the results for different $n_s$. Fig.~3 shows the bulk 
quasi-particle weight $z$ as a function of $U$ as obtained from
calculations for $n_s=8$ and $n_s=10$ sites. We notice that there
is a good agreement between both results for $U$ smaller than
$\sim 14.5$ or $z$ larger than $\sim 0.01$. For $z < 0.01$ the
energy scale set by the width of the quasi-particle peak 
$\approx z W$ can no longer be resolved, i.~e.\ $\Delta E \approx
0.12=W/100$. For $n_s=8$ and $U=15$ this $\Delta E$ is also the 
typical energetic distance between the conduction-band level at 
the Fermi energy, $\epsilon_k=0$, and the level with lowest 
absolute energy $\epsilon_{k'} \ne 0$ in the self-consistent 
solution. The error that is introduced by the low-energy 
cutoff $\widetilde{\beta}$ is found to be of minor importance 
compared with the error due to finite-size effects. This is 
ensured by choosing $\widetilde{\beta}$ such that 
$\pi \widetilde{\beta}^{-1}$ ($\approx 0.03$) 
is smaller than $\Delta E$. 

From the zero of $z(U)$ in Fig.~3 we can estimate the critical 
interaction: $U_{c2} \approx 16 t = 1.33 W$ for $n_s=8$ and 
$U_{c2} \approx 15t = 1.25 W$ for $n_s = 10$. Obviously, a precise 
determination is not possible. The results may be compared with 
the value $U_{c2} \approx 1.34 W$ which has been obtained for 
the Bethe lattice with infinite connectivity using ED at a finite 
but low temperature $\beta t=100$ \cite{LGK94}. For interaction
strengths $U > U_{c1} = 11.5 t$ we also obtain an insulating 
solution. Contrary to $U_{c2}$, the value found for 
$U_{c1}$ is almost independent of $n_s$. This is plausible since 
there is no small energy scale in the case of the insulator.

We now return to our main subject of interest: the layer 
and interaction dependence of the quasi-particle weight at 
the surfaces of the s.c.\ lattice. Fig.~4 shows the results 
obtained for $n_s=8$. At weak coupling we notice a quadratic
$U$ dependence consistent with perturbation theory. For the 
sc(100) surface there are no significant differences between 
the DMFT and the SOPT result for $z_\alpha$ up to $U \approx 2-3$ 
(see top panel in Fig.~4). The same holds for the sc(110) surface
(not shown). Contrary, perturbation theory breaks down at a much 
weaker interaction strength for the sc(111) surface as it is 
obvious comparing the results in Figs.~2 and 3.

For low and intermediate $U$, $z_\alpha$ has an oscillating layer 
dependence while it is monotonous close to the transition (Fig.~4, 
insets). The strongest surface effects are seen for the open sc(111) 
surface while for the sc(100) surface $z_\alpha \approx z_{\rm bulk}$ 
except for $\alpha = 1$. This trend is related to the decrease of 
the surface coordination numbers: $n_{\rm S}^{(100)}=5$, 
$n_{\rm S}^{(110)}=4$, $n_{\rm S}^{(111)}=3$, to be compared 
with the bulk value $n_{\rm B}=6$.

For all surfaces and layers there is a unique and common critical 
interaction $U_{c2}$ where $z_\alpha(U)$ approaches zero. The 
value of $U_{c2}$ is the same as the bulk critical interaction
strength (found for $n_s=8$), i.~e.\ $U_{c2}$ is determined by 
the bulk system.
In all cases the quasi-particle weight of the topmost layer 
$z_{\rm S}$ is significantly reduced with respect to the bulk 
value $z$. This is plausible since the variance $\Delta = 
n_{\rm S} t^2$ of the free ($U=0$) surface density of states 
is smaller due to the reduced surface coordination number 
$n_{\rm S}$, and thus $U/\sqrt{\Delta}$ is larger at the surface 
which enhances correlation effects. The reduced surface coordination 
number therefore {\em tends} to drive the surface to an insulating 
phase at an interaction strength lower than the bulk critical 
interaction $U_{c2}$ ($z > 0$, $z_{\rm S} = 0$). A real surface 
transition, however, is not found; $z_{\rm S}$ remains to
be non-zero up to $U_{c2}$. 

%++++++++++++++++++++++++++++++++++++++++++++++++++++++++++
\begin{figure}[t] 
\vspace{-2mm}
\centerline{\psfig{figure=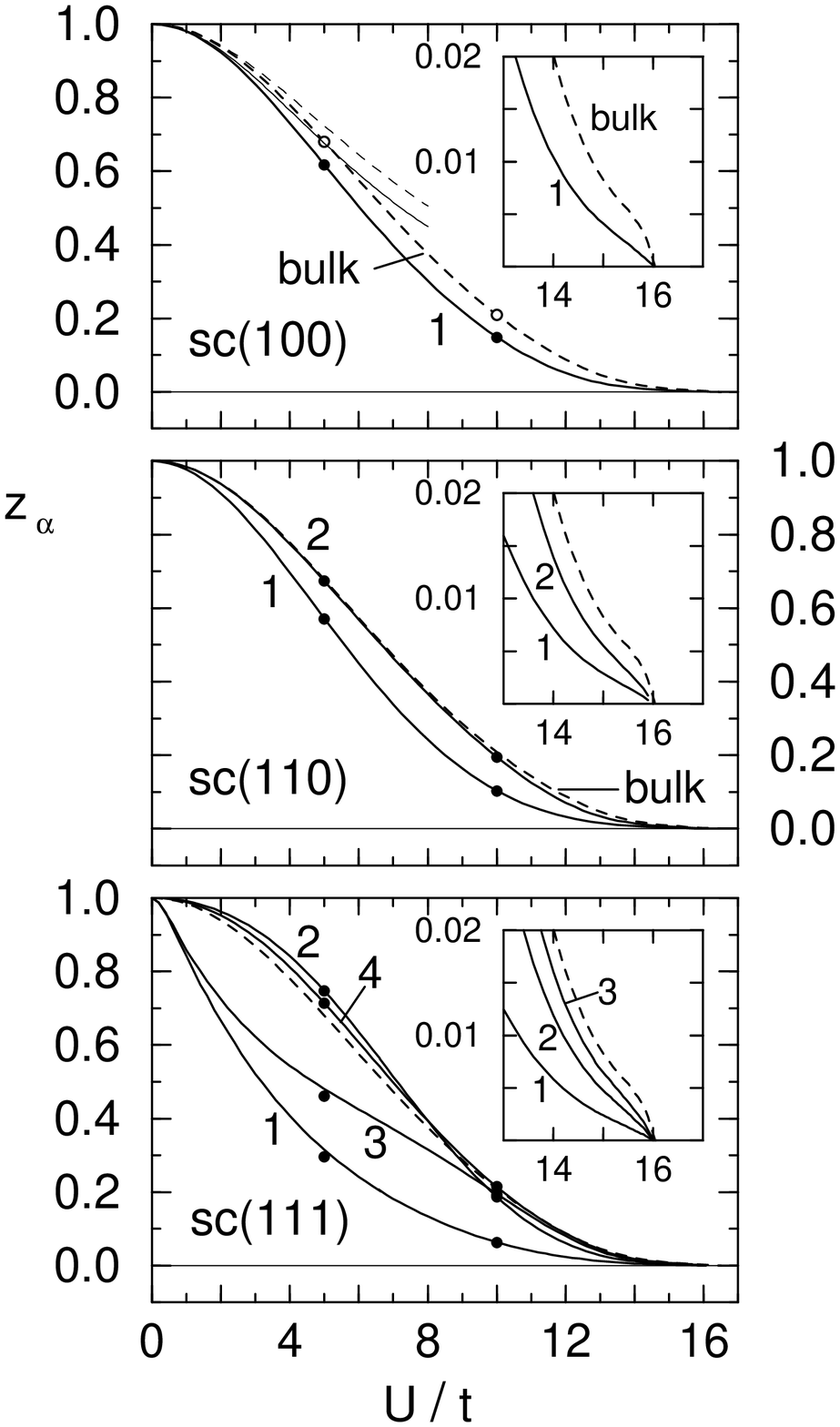,width=100mm,angle=0}}
\vspace{-2mm}

\parbox[]{85mm}{\small Fig.~4.
$U$ and layer dependence of 
$z_\alpha=(1-\beta_\alpha)^{-1}$ in the vicinity of the surface
(bulk: dashed lines).
ED calculation for 8 sites (filled/open circles: $n_s=10$).
$N_\perp=11, 15, 21$ for the sc(100), sc(110), sc(111) surface, 
respectively. Insets: $z_\alpha$ in the critical regime.
Thin (solid and dashed) lines: SOPT-HF results for the sc(100) 
surface and $U<8$ (top panel).
\label{fig:zofu}
}
\end{figure}
%++++++++++++++++++++++++++++++++++++++++++++++++++++++++++

This is interpreted as follows: 
For $U < U_{c2}$ the bulk quasi-particle band has a finite 
dispersion. The corresponding low-energy excitations are thus 
extended over the entire lattice. Due to hopping processes between 
the very surface and the bulk, they will have a finite weight also 
in the top layer. Therefore, below $U_{c2}$ the bulk excitations 
will to some extent {\em induce} a quasi-particle peak with a 
non-zero weight $z_{\rm S} > 0$ in the topmost layer. To test this 
interpretation, we decoupled the top layer from the rest system by 
switching off the hopping between the top and the subsurface layer, 
$t_{12}=0$. For the sc(111) surface this implies $n_{\rm S}=0$. 
The top-layer self-energy is thus given by the (insulating) 
atomic-limit expression $\Sigma_{\alpha=1}(E)=U/2+U^2/4E$, while 
for the other layers we get the same layer-dependent (metallic) 
self-energy as before but shifted by one layer. Taking this as 
the starting point for the DMFT self-consistency cycle where
$t_{12}=t$ again, we observe that a finite quasi-particle weight 
in the top layer $z_{\rm S}>0$ is generated immediately and that 
the cycle converges to the same solution as before.

There is no Mott localization at the surface of the metallic 
Fermi liquid. However, the tendency of the system to reduce the 
surface quasi-particle weight (enhance the surface effective mass) 
results in a partial (wave-vector-dependent) localization 
of quasi-particles at the surface: Let us consider the coherent 
part of the surface electronic structure. To make the calculations 
more transparent, we assume that $z_\alpha=z$ except for the top 
layer ($z_{\alpha=1}=z_{\rm S}$). 
This is well justified for the sc(100) 
and, except for the very critical regime, also for the sc(110) 
surface (Fig.~4). Furthermore, we define the ratio $r^2=z_{\rm S}/z$. 
Two-dimensional Fourier transformation of the renormalized hopping 
matrix ${\bf \overline{T}} = \sqrt{\bf Z} {\bf T} \sqrt{\bf Z}$ 
yields a tridiagonal matrix $\overline{T}_{\alpha \alpha'}({\bf k})$ 
in the layer indices $\alpha,\alpha'=1 , ... , N_\perp$. For 
$\alpha,\alpha' \ge 2$ the non-zero elements are given by 
$\overline{T}_{\alpha \alpha} ({\bf k}) = z \epsilon_\|({\bf k})$ 
and $\overline{T}_{\alpha \alpha \pm 1} ({\bf k}) = 
z \epsilon_\perp({\bf k})$. For the sc(100) surface the parallel 
and perpendicular dispersions read \cite{other}: 
$\epsilon_\|({\bf k}) = 2 t (\cos(k_xa) + \cos(k_ya))$ and
$\epsilon_\perp({\bf k}) = t$. At the very surface we have:
$\overline{T}_{11} ({\bf k}) = r^2 z \epsilon_\|({\bf k})$
and $\overline{T}_{12/21} ({\bf k}) = r z \epsilon_\perp({\bf k})$.
The specific surface renormalization ($r^2<1$) leads to two 
(localization) effects relevant on the energy scale $z W$: 
a diminished parallel dispersion in the top layer and a tendency 
to decouple the top-layer from the bulk spectrum.

%++++++++++++++++++++++++++++++++++++++++++++++++++++++++++
\begin{figure}[t] 
\vspace{-2mm}
\psfig{figure=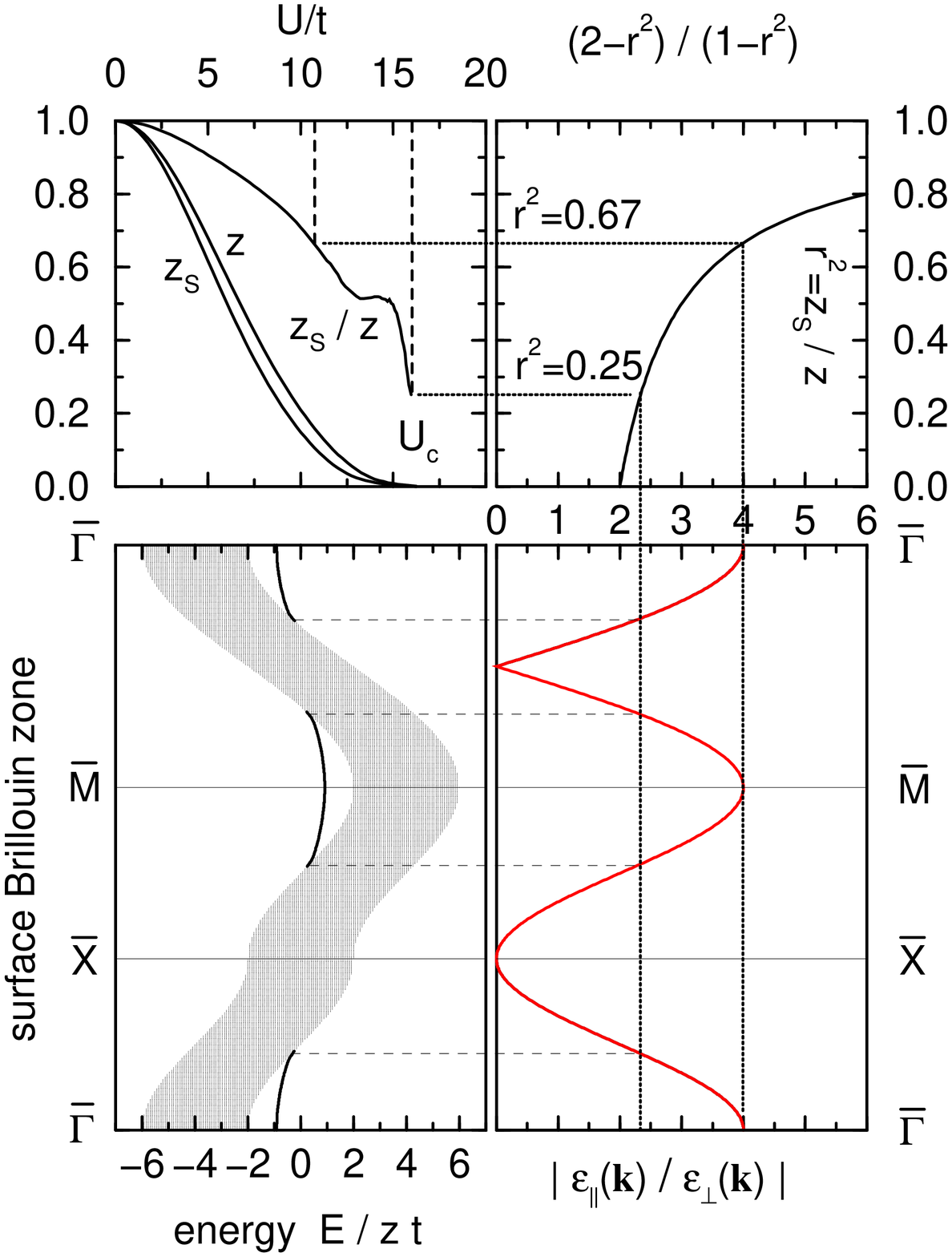,width=80mm,angle=0}
\vspace{0mm}

\parbox[]{85mm}{\small Fig.~5.
{\em Upper left:} $z(U)$, $z_{\rm S}(U)$ and the ratio 
$r^2=z_{\rm S}/z$ for the sc(100) surface.
{\em Upper right:} $(2-r^2)/(1-r^2)$ as a function of $r^2$. 
{\em Lower right:} Ratio between parallel and perpendicular 
dispersion along high-symmetry directions in the two-dimensional
surface Brillouin zone (SBZ). At ${\bf k}$ points with 
$(2-r^2)/(1-r^2) <
|{\epsilon_\|({\bf k})}/{\epsilon_\perp({\bf k})}|$ surface 
excitations are split off the bulk continuum.
{\em Lower left:} Coherent part of the quasi-particle spectrum
projected onto the SBZ calculated from Eqs.\ (\ref{eq:g1}) and
(\ref{eq:g2}). 
Filled area: bulk continuum. Solid line: surface excitation 
for $r^2=0.25$ corresponding to $U=U_{c2}$.
\label{fig:exist100.eps}
}
\end{figure}
%++++++++++++++++++++++++++++++++++++++++++++++++++++++++++

As a consequence, surface excitations split off the continuum of
bulk excitations in the SBZ: 
For tridiagonal $\overline{T}_{\alpha \alpha'} ({\bf k})$ the 
$\bf k$-resolved surface Green function is readily calculated 
\cite{Hay80}; we have:
\begin{eqnarray}
  G_0({\bf k},E) = z \frac{1}{2b^2} 
  \big(E-a \mp \sqrt{(E-a)^2 - 4 b^2} \big)
  \nonumber \\ (\mbox{for} \;\; \pm \mbox{Re} (E -a )> 0) \: ,
\label{eq:g1}
\end{eqnarray}
with $a \equiv z \epsilon_\|({\bf k})$, 
$b \equiv z \epsilon_\perp({\bf k})$
for the unperturbed surface ($r^2=1$). Thus, for $r^2=1$ 
the coherent part of the surface spectral density
$-\mbox{Im}G_0({\bf k},E+i0^+)/\pi$ is semi-elliptical for each 
$\bf k$ with band edges at $E = a \pm 2b$. The surface Green 
function $G$ for the perturbed surface ($r^2<1$) can be expressed 
in terms of $G_0$ as \cite{Hay80}:
\begin{equation}
  G({\bf k},E) = \frac{z_{\rm S}}{ 
  E-r^2 a -
  r^2 b^2 G_0({\bf k},E)
  }  \: .
\label{eq:g2}
\end{equation}

A surface excitation is characterized by an energy that (for a given
wave vector ${\bf k}$ of the SBZ) falls outside the continuum of
bulk excitations described by $G_0$. A condition for the existence 
of surface excitations is therefore given by $|E_{\rm S}-a| > 2b$ 
where $E_{\rm S}$ is the excitation energy obtained from the 
perturbed surface Green function: $G({\bf k},E=E_{\rm S})^{-1} = 0$. 
With $G_0({\bf k},E) = \pm z/b$ at the band edges $E = a \pm 2b$
this is equivalent to:
\begin{equation}
  0 > 2 b^2 \pm b (a - a_0) - b_0^2 \: ,
\label{eq:surf}
\end{equation}
where $a_0 \equiv r^2 z \epsilon_\|({\bf k}) = r^2 a$ and
$b_0 \equiv r z \epsilon_\perp({\bf k}) = r b$. 
Eq.\ (\ref{eq:surf}) generally tells us whether or not a discrete 
eigenvalue is split off the eigenvalue continuum of a semi-infinite
tridiagonal matrix with modified ``surface'' parameters.
For the present case it immediately leads to the following 
simple criterion for the existence of a surface excitation:
\begin{equation}
  \frac{2-r^2}{1-r^2} < 
  \left| \frac{\epsilon_\|({\bf k})}
  {\epsilon_\perp({\bf k})} \right| \: .
\label{eq:ss}
\end{equation}
Hereby, the ratio $r^2=z_{\rm S}/z$ is related to the ratio between 
the free dispersions.

Fig.~5 visualizes the criterion for the case of the sc(100) surface.
At small $U$, the renormalization factor $r^2$ is too close to 1
to generate surface excitations. These appear above $U\approx 11t$
and split off the bulk continuum at the $\overline{\Gamma}$ and
$\overline{M}$ high-symmetry points in the SBZ. The minimum
interaction strength is well below $U_{c2}$. Finite-size effects
are of no importance here. For $U\mapsto U_{c2}$
surface excitations are found in a fairly extended region around 
$\overline{\Gamma}$ and $\overline{M}$. In the insulating phase 
the whole coherent spectrum diappears. The sc(110) surface is 
even more 
favorable for surface excitations. Eq.\ (\ref{eq:ss}) always holds 
along the $k_y=\pi/\sqrt{2}a$ direction 
where $\epsilon_\perp({\bf k}) = 0$ \cite{other}.
For the sc(111) surface $z_\alpha$ deviates from $z$ also in some 
subsurface layers. Calculating the coherent surface spectrum
from the DMFT results for $z_\alpha(U)$ shows that surface
excitations do not exist for any $U$.

As can be seen in Fig.~5, the surface excitation splits off the bulk
continuum away from the Fermi energy. Due to quasi-particle damping 
the surface peak in the spectrum will therefore aquire a finite width 
which is of the order $\Delta E \approx z_{\rm S} \cdot|\gamma_{\rm S}| 
E_{\rm S}^2$ where $\gamma_{\rm S}$ is the damping coefficient at 
the surface defined in Eq.\ (\ref{eq:sigmaexp}). Since we have
$E_{\rm S} \approx z_{\rm S} t$ for the position of the excitation,
the width is $\Delta E \approx z_{\rm S}^3 |\gamma_{\rm S}| t^2$. For 
$U \mapsto U_{c2}$ ($z\mapsto 0$) the $E^2$ coefficient in the 
expansion of the self-energy diverges as $\gamma \sim -1/z^2 W$ 
as has been shown in Ref.\ \cite{MSKR95} for the $d=\infty$ Bethe 
lattice. This result may be used to estimate 
$\Delta E < z_{\rm S} t$. 
Comparing $\Delta E$ with the energetic distance between 
the surface peak and the edge of the bulk continuum, 
$|E_{\rm S} - E_{\rm B}| \approx z t$ (Fig.~5), shows both
quantities to be of the same order of magnitude 
for $U \mapsto U_{c2}$. We conclude that the surface peak 
cannot be fully separated in energy from the bulk features
in the surface-projected spectrum.

A complete energetic separation between bulk and surface excitations 
is possible if $E_{\rm S}$ crosses the Fermi energy as a function 
of ${\bf k}$. The dispersion of the surface mode can be calculated 
from (\ref{eq:g1}) and (\ref{eq:g2}) to be:
\begin{equation}
  E_{\rm S}({\bf k}) = \frac{1}{2} z r^2 \epsilon_\|({\bf k}) \pm
  z r^2 \sqrt{\frac{\epsilon_\|({\bf k})^2}{4} -
  \frac{\epsilon_\perp({\bf k})^2}{1-r^2} } \: .
\end{equation}
Setting $E_{\rm S}({\bf k}) = 0$ yields $\epsilon_\perp({\bf k}) = 0$.
As mentioned above, this is fulfilled e.~g.\ along the 
$k_y=\pi/\sqrt{2}a$ direction for the sc(110) surface.
\\

{\center \bf \noindent VI. CONCLUSION \\ \mbox{} \\} 

While the investigation of the {\em surface} electronic structure
of a metal has a long tradition, there are only few attempts to 
account for correlation effects beyond the Hartree-Fock
approximation or the (ab initio) local-density approach (see 
\cite{PN97c,mag} and references therein). The present paper has 
focussed on the Hubbard model as a standard model for correlated
itinerant electrons and on the Mott transition as a genuine 
correlation effect. Adapting the dynamical mean-field theory for
general film geometries and using the exact-diagonalization approach, 
we have investigated the specific surface properties of the $T=0$ 
semi-infinite Hubbard model at half-filling.

At the surface electron-correlation effects are found to be 
significantly more pronounced compared with the bulk. The top-layer 
quasi-particle weight is considerably reduced for all surfaces and 
interaction strengths. Surface correlation effects are the stronger 
the larger is the reduction of the surface coordination number. For 
the comparatively open sc(111) surface there is a pronounced
layer dependence of the quasi-particle weight $z_\alpha$ for 
weak and intermediate couplings. As $U$ approaches the critical
region, the layer dependence becomes monotonous in all cases. 

The reduced surface coordination number {\em tends} to drive 
the system to a phase with an insulating surface on top of a 
metallic bulk. Eventually, however, the low-energy bulk excitations 
{\em induce} a finite spectral weight at the Fermi energy in the
top-layer density of states. Consequently, all layer-dependent 
quasi-particle weights vanish at the same coupling strength 
$U_{c2}$; there is a unique transition point. This excludes 
a modified surface critical interaction and thus the existence 
of a surface phase. 

On the other hand, a different localization effect has been found: 
There are special regions in the surface Brillouin zone, where the 
low-energy excitations in the top layer cannot couple to the bulk 
modes and are confined to two-dimensional lateral propagation. 
These surface excitations are of particular interest since their 
nature is rather different compared with the well-known Tamm- and 
Shockley-type surface states \cite{Tam32,Sho39}. Opposed to the 
latter they are caused by electron correlations exclusively. Quite
generally, as long as the concept of a highly renormalized Fermi 
liquid applies, the specific surface renormalization of the 
effective mass tends to generate excitations split off the bulk 
continuum. This mechanism is sufficiently general to survive also 
at finite temperatures and non-bipartite lattices where 
antiferromagnetic order is expected to be suppressed. High-resolution 
photoemission from single-crystal samples of transition-metal oxides 
should be appropriate to detect the excitation by suitably tuning 
the Mott transition.

The present study has been restricted to surface geometries and 
uniform model parameters. It is an open question whether or not 
there is a surface phase if the hopping or the interaction
strength at the very surface is modified. Possibly, a metallic
surface coexisting with a Mott-insulating bulk can be found for
a strongly reduced surface $U$. Future investigations may also 
concern thin-film geometries where the study of the thickness 
dependence of the critical interaction is of particular interest.
\\

{\center \bf \noindent Acknowledgement \\ \mbox{} \\} 

This work is supported by the Deutsche Forschungsgemeinschaft 
within the SFB 290.
\vspace{5mm}

---------------------------------------------------------------------

\end{document}